\documentclass{article}
\usepackage[preprint]{neurips_2019} 
\setcitestyle{numbers}

\usepackage[utf8]{inputenc} 
\usepackage[T1]{fontenc}    
\usepackage{hyperref}       
\usepackage{url}            
\usepackage{booktabs}       
\usepackage{amsfonts}       
\usepackage{nicefrac}       
\usepackage{microtype}      
\usepackage{amsmath}

\usepackage{graphicx}
\usepackage{caption}
\usepackage{subcaption}
\usepackage{bm}
\usepackage{MnSymbol}
\usepackage{wasysym}
\usepackage[toc,page]{appendix}

\title{Linking Social Media Posts to News with Siamese Transformers}

\author{%
  Jacob Danovitch \\
  Institute for Data Science\\
  Carleton University\\
  Ottawa, CA \\
  \url{jacob.danovitch@carleton.ca} \\
}

\DeclareMathOperator{\att}{Attention}
\DeclareMathOperator{\mhatt}{MultiHeadAttention}
\DeclareMathOperator{\softmax}{softmax}
\DeclareMathOperator{\expsum}{expsum}

\renewcommand{\H}{\bm{H}}
\newcommand{\seq}[1]{\left[\bm{#1}\right]}
\newcommand{\ev}[1]{\mathbb{E}\left[#1\right]}
\newcommand{\norm}[1]{\bm{\lVert#1\rVert}}

\begin{document}

\maketitle

\begin{abstract}
Many computational social science projects examine online discourse surrounding a specific trending topic. These works often involve the acquisition of large-scale corpora relevant to the event in question to analyze aspects of the response to the event. Keyword searches present a precision-recall trade-off and crowd-sourced annotations, while effective, are costly. This work aims to enable automatic and accurate ad-hoc retrieval of comments discussing a trending topic from a large corpus, using only a handful of seed news articles.
\end{abstract}

\section{Introduction}
Many computational social science projects examine online discourse surrounding a specific trending topic, such as political events, natural disasters, or sporting matches. These works often involve the acquisition of large-scale corpora relevant to the event in question to analyze aspects of the response to the event, often with a focus toward social media. However, existing methods for the acquisition of these events come with trade-offs. 

Commonly, keyword searches are used \cite{mazoyer2018real}, which present a choice between high precision (using narrow keywords and getting fewer but more accurate results) and recall (using broad keywords and getting many irrelevant results). One common method of keyword searching, specific to Twitter, is the use of hashtags. However, hashtag-only searches can lead to low recall \cite{letierce2010understanding}. Crowd sourcing is an effective way to filter corpora \cite{lin2016linking}, though it can be expensive both monetarily as well as in terms of time spent. This work proposes a method for automatic and accurate ad-hoc retrieval of comments discussing a trending topic from a large corpus using only a handful of seed news articles. Using a Siamese architecture and triplet loss, we jointly embed news articles and social media posts with the objective of minimizing the distance between the embeddings of each article and its relevant posts. This allows the automatic filtering of corpora by selecting comments most similar to news articles describing the event or topic of interest. The asymmetric lengths of comments and news articles pose a challenge for Siamese architectures. We propose a novel solution using a sparse attention mechanism, which allows the network to attend only to the most relevant parts of the input, easing the asymmetry in length. We make our code publicly available\footnote{\url{https://github.com/jacobdanovitch/jdnlp}}.

\section{Related Work}
The specific task of matching articles and social media comments (used interchangeably with \textit{posts}) has received limited attention, especially in recent years. Some early works focused on \textit{social content alignment}, the task of aligning comments to specific sentences within articles. Latent Dirichlet Allocation \cite{Sil2011ReadAlongRA, hoang2014bridging} and several extensions, such as the Document-Comment Topic Model \cite{socialalign, hou2017learning} and Specific-Correspondance LDA \cite{Das:2014:GBC:2556195.2556231}, were used to provide interpretable alignments of comments to the most relevant segments of a news article. However, binary classification with feature engineering was seen to outperform these methods \cite{Sil2011SupervisedMO}.

Similar methods have been employed for the task of matching tweets to news articles. \cite{guo2013linking} proposes a graphical weighted matrix factorization approach while also contributing a large-scale dataset similar to ours. Recent work using this dataset proposes an interactive attention mechanism to match tweet-article pairs \cite{zhao2019interactive}. These works find the most relevant news article for a particular tweet, rather than vice versa. The intent of our work is more similar to \cite{doi:10.1177/0165551516653082}, which identifies the most relevant tweet for a given news article using binary classifiers as well as semantic similarity.

Little has been done to apply more recent advancements in deep learning to this task. This is likely due to the large computational overhead in training over corpora of news articles, which consist of thousands of tokens each, as well as the inherent theoretical challenges these methods face in processing long documents. These challenges have inspired creative solutions, such as segmenting long documents into individual sentences, hierarchically processing each sentence, and performing an aggregation step \cite{Chen_2019}. Alternatively, \cite{liu2018matching} presents an approach using Graph Convolution Networks on \textit{concept-interaction graphs} to match pairs of news articles. While this is an effective method, the approach is restricted to pairwise classification, which creates a combinatorial explosion of pairs. Our work uses a Siamese architecture to perform retrieval using only cosine similarity.




\section{Methodology}
\subsection{Task definition}

We use a large corpus of tweets which share links to news articles published by CNN and the New York Times \cite{guo2013linking}. The corpus contains $34,888$ tweets which link to $12,704$ unique articles. While the goal of the original study is to find the most relevant news article for a given tweet, we are interested in the reverse; finding the most relevant tweets for a given news article. We say that for each tweet-article pair $(T, A)$, $T$ is \textit{relevant} to $A$, and any other tweet $T'$ not paired with $A$ is \textit{irrelevant}.

\subsection{Notation}

We refer to vectors in bold ($\mathbf{x}$), tensors in capitalized bold ($\mathbf{X}$), and scalars in standard font ($x,X$). We define an encoder as a deep neural network which maps a sequence of $S$ word embeddings $\seq{e_1, e_2, \ldots, e_S}$ to either an output sequence of $D$-dimensional hidden states $\H = \seq{h_1, h_2, \ldots, h_S}$ (in the sequence-to-sequence case) or a single $D$-dimensional representation $\H$ (in the sequence-to-vector case).

\subsection{Objectives}

We wish to use a Siamese network to jointly learn fixed-length embeddings for each $(T, A)$ pair, minimizing the distance between them using triplet loss. Projecting $A$ and $T$ into a common latent space allows us to compare the similarity of documents from vastly different domains (short, informal tweets versus long, formal news articles). However, doing so requires an encoder capable of solving several challenges pertaining to document length.

\paragraph{Noise.} News articles can be in the order of thousands of tokens, and they often discuss multiple different topics \cite{hou2017learning}. The asymmetric length of tweets relative to the articles suggests that tweets rarely discuss the entirety of the article's content, making much of the article irrelevant and perhaps even adding additional noise. As such, it would be useful for an encoder to be able to filter out this noise and focus on only the most important input tokens. 

\paragraph{Efficiency.} Using large batch sizes creates more informative triplets \cite{Wu_2017}, which is constrained by the already-high memory usage of processing thousands of tokens at once. These concerns are accentuated by the computational demands of the attention mechanism. While attention is more effective than other methods for modelling long sequences, its $\mathcal{O}(n^2)$ complexity \cite{vaswani2017attention} will exhaust reasonable resources with such long documents. As such, we require an encoder capable of efficiently processing long documents without sacrificing quality in the process.


\subsection{Blocking out the noise}

We begin by confronting the theoretical challenges of modelling long documents, such as the ability to capture the long-range dependencies and focusing only on crucially important tokens. Attention-based architectures have shown to outperform alternative candidates for these challenges \cite{raffel2015feedforward}. However, certain models such as BERT \cite{devlin2018bert} can be computationally intensive and impose maximum token lengths, which prevents their use. An additional problem for attention-based models introduced by such long documents is that they may struggle to identify vital portions of the text. For example, consider the attention mechanism and its multi-headed extension \cite{vaswani2017attention}: 


\begin{align}
    \att(\mathbf{Q, K, V}) &= \softmax(f(\mathbf{Q, K))V} \\
    \bm{a_i} &= \att(\bm{QW}_{\bm{Q}_i}, \bm{KW}_{\bm{K}_i}, \bm{VW}_{\bm{V}_i}) \\ 
    \mhatt(\mathbf{Q, K, V}) &= \seq{a_1; a_2; \ldots a_h}
\end{align}


Where the parameters $\bm{W}_{\bm{Q}_i}, \bm{W}_{\bm{K}_i}, \bm{W}_{\bm{V}_i} \in \mathbb{R}^D$ are trainable and $f$ is a function which computes the similarity between each entry in $\bm{Q}$ and $\bm{K}$. One such function is $\cos(\mathbf{Q, K}) = \mathbf{\frac{QK^T}{\norm{Q} \cdot \norm{K^T}}}$, which is known as cosine attention \cite{graves2014neural}. Crucial to the attention operation is the $\softmax$ function:

\begin{equation}
    \sigma(\bm{z})_i = \frac{e^{z_i}}{\sum^{S}_{j=1} e^{z_j}}
\end{equation}

This produces an $S$-length vector with a sum of $1$, where each $z_i$ is a weight for each token in the input document. The operation $\softmax(\cdot)\bm{V}$ is equivalent to taking a weighted average of each token embedding in $\bm{V}$. Note that $\softmax$ has full support (all $\sigma(\bm{z})_i > 0$), which means that even unimportant portions of the input document will still receive non-zero weight \cite{Correia_2019}. As we noted earlier, our documents are very long and many of the words could be unimportant to the output, which makes it highly difficult for $\softmax$ to filter them out without assigning any weights of precisely $0$. In fact, we observe a negative correlation between $S$ and $\ev{\sigma(\bm{z})_i}$ - that is, as documents grow \textit{longer}, the expected weight of a given token \textit{decreases}. We provide a brief proof in appendix \ref{appendix:expsum}.

\begin{figure}[hbtp]
    \centering
    \includegraphics[width=0.8\linewidth]{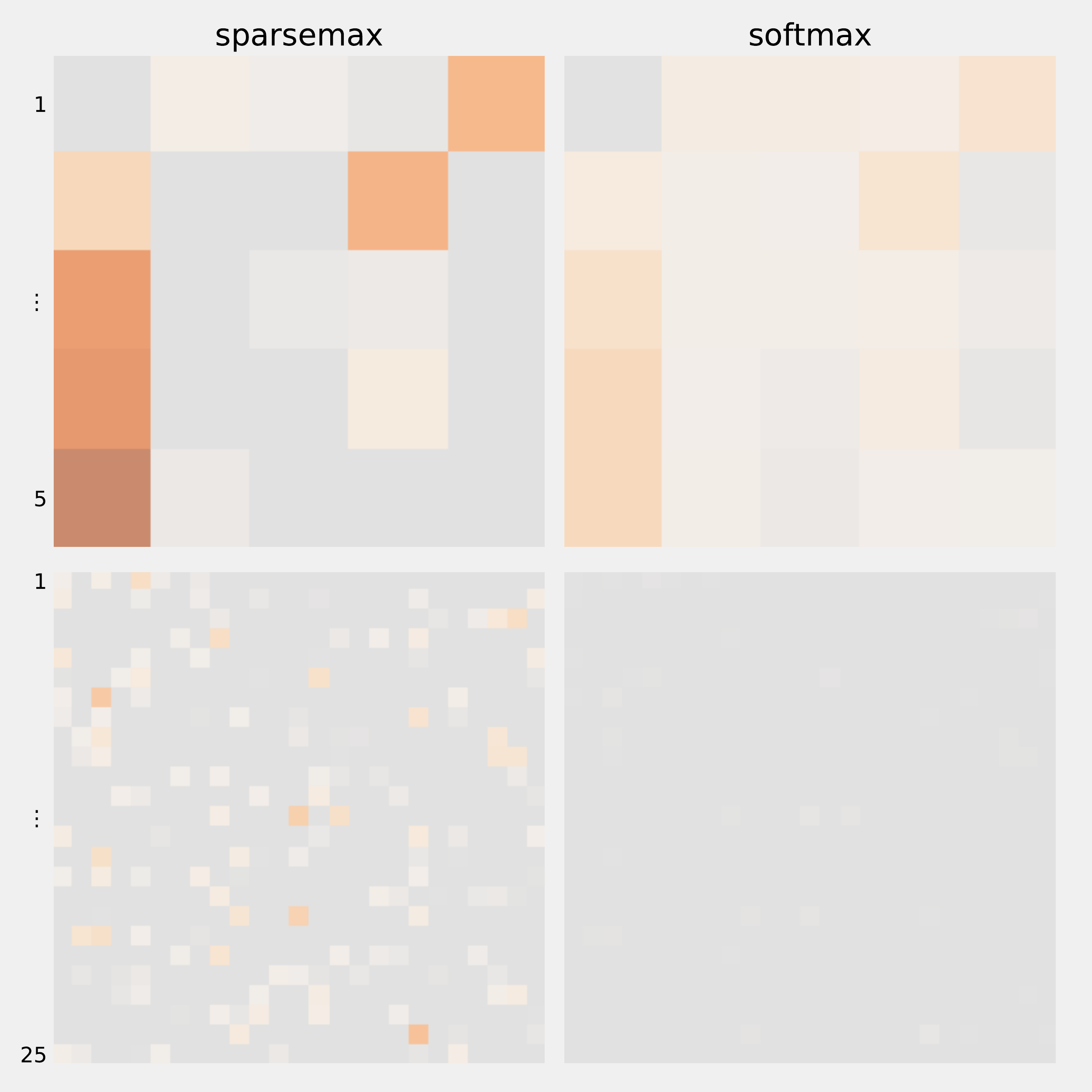}
    \caption{Self-attention applied to sequences of 5 and 25 random embeddings using $\softmax$ and sparsemax. As document length increases, $\softmax$ produces nearly uniform weights approaching $0$, making self-attention equivalent to taking the mean over all embeddings.}
    \label{fig:attn_comp}
\end{figure}

This is problematic, as it will be difficult for the most important tokens to stand out in such long sequences like news articles. As the value of each attention weight decreases, the distribution becomes more uniform, and the operation becomes equivalent to simply taking the mean over all $\bm{h_i} \in \H$, many of which should have been filtered out. To counter this problem, we look toward the growing family of sparse activation functions. Figure \ref{fig:attn_comp} demonstrates a comparison of self attention using $\softmax$ versus one such sparse function, known as sparsemax \cite{martins2016softmax}. While both activations identify important tokens in a sequence of $5$, $\softmax$ approaches uniformity for the sequence of length $25$ while sparsemax still identifies important tokens.

In particular, we replace $\softmax$ with the $\bm{\alpha}-\mathrm{entmax}$ activation function, a controllably sparse alternative \cite{peters-etal-2019-sparse}, and we use a unique trainable parameter $\bm{\alpha}_h$ for each attention head $h$. This is known as adaptively sparse multi-head attention, which addresses our first objective by allowing each attention head to learn an optimal level of sparsity \cite{Correia_2019}. This allows each head to learn which segments of the input sequence to filter, reducing the noise introduced by long input documents.

\subsection{Improving efficiency}

Next, we address our second objective, which requests an encoder capable of efficiently processing long documents without sacrificing quality in the process. Irrespective of the activation function used, the $\mathcal{O}(n^2)$ complexity of a standard Transformer layer is prohibitively expensive for documents thousands of tokens long. As such, we require a way to model long-range dependencies more efficiently than a standard Transformer. For this, we use the Star Transformer, a simple and efficient extension of the standard Transformer capable of learning many of the same semantic relationships and long-range dependencies \cite{guo2019startransformer}. 

The Star Transformer is able to reduce the quadratic time complexity to linear time ($\mathcal{O}(n)$) by using a message-passing mechanism along a sliding window of embeddings. Given a sequence of input embeddings $\left[\mathbf{e}_1, \mathbf{e}_2, \ldots, \mathbf{e}_n\right]$, we produce an output sequence of embeddings $\seq{\mathbf{h}_1, \mathbf{h}_2, \ldots, \mathbf{h}_n}$. Initially, each $\mathbf{h}_i = \mathbf{e}_i$, and a message-passing relay node $\mathbf{s}$ is initialized as the mean of all $\mathbf{e}_i$. For each embedding $\mathbf{e}_i$, its immediate $c$ neighbors and the relay node $\mathbf{s}$ are used to form a new representation $\mathbf{h}_i$. This message-passing mechanism enables the learning of long-range dependencies in linear time, as each attention operation only considers the $c$ immediate neighbors of each token (as well as the relay node).

\begin{align}
    \mathbf{C}_i &:= \left[\mathbf{h}_{i-c}; \ldots \mathbf{h}_{i}; \ldots \mathbf{h}_{i+c}; \mathbf{s}\right] \\
    \mathbf{h}_i &:= \mhatt(\mathbf{C}_i, \mathbf{h}_i, \mathbf{h}_i) 
\end{align}

After all $i$ operations, we then update the relay node as $\mathbf{s} := \mhatt(\mathbf{H, s, s})$ where $\mathbf{H} = \seq{\mathbf{h}_1; \mathbf{h}_2; \ldots \mathbf{h}_n}$. The authors find that performing $T$ iterations of these operations enables re-reading to more effectively capture long-range dependencies, though additional iterations only marginally improved performance in our experiments. To encode the output sequence $\mathbf{H}$ to a fixed-length vector, we take the maximum over each embedding dimension and average it with the final state of the relay node, computed as $\frac{\max(\mathbf{H})+\mathbf{s}}{2}$.

\subsection{Siamese Architecture}

Finally, we use our altered Star Transformer to jointly embed each tweet-article pair, represented as a tuple of word embedding sequences $(\left[\bm{t}_1, \bm{t}_2, \ldots, \bm{t}_m\right], \left[\bm{a}_1, \bm{a}_2, \ldots, \bm{a}_n\right])$. Each are independently passed through the encoder, producing fixed-length vectors $\bm{f_t}$ and $\bm{f_a}$. These embeddings are directly passed to the objective function, triplet loss \cite{chechik2010large}, formulated as:

\begin{equation}
    L = \max(0, \norm{f_a - f_t}^2 - \norm{f_a - f_{t'}}^2 + \alpha)
\end{equation}

\begin{figure}
    \centering
    \includegraphics[width=0.85\linewidth]{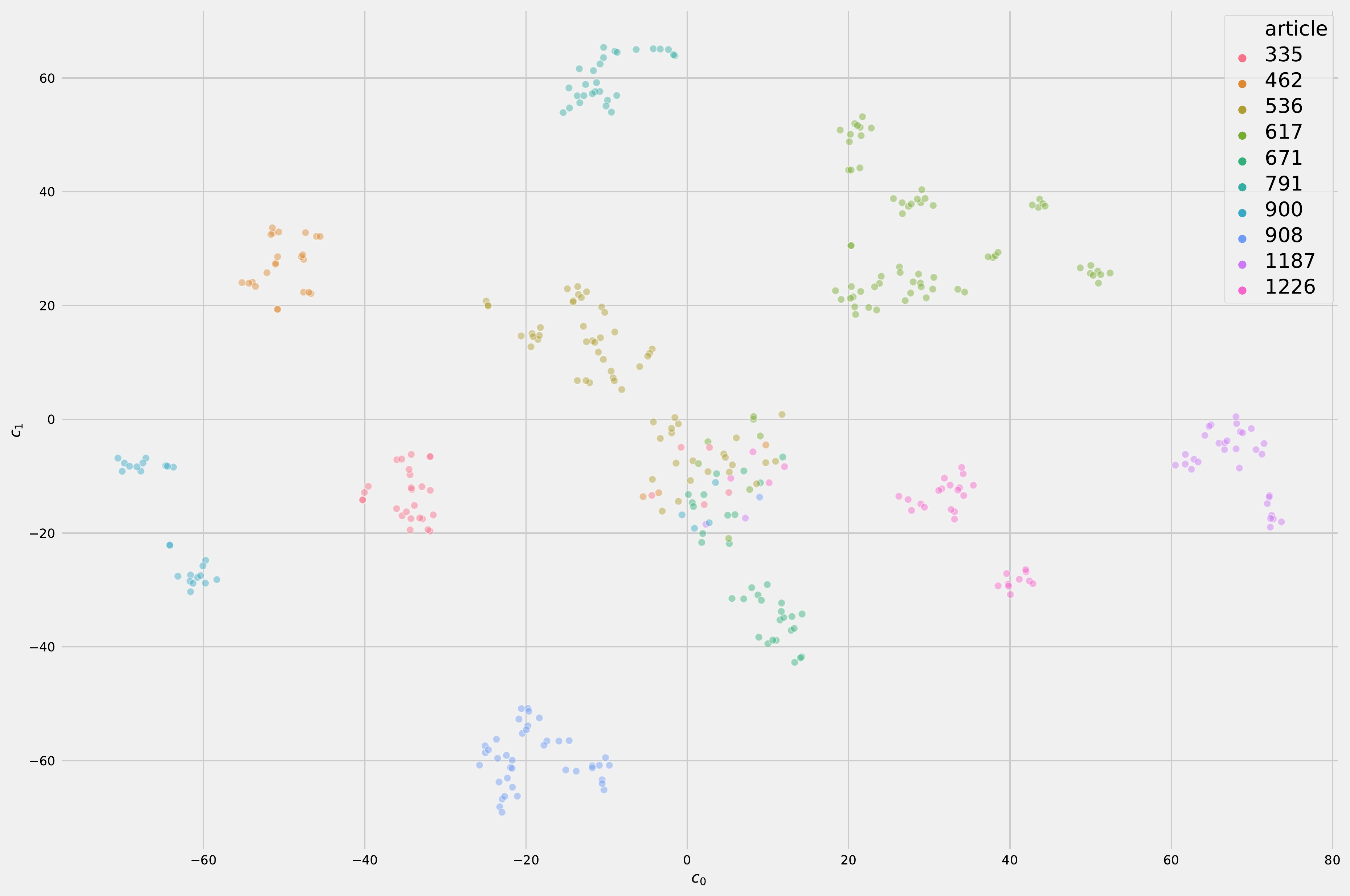}
    \caption{t-SNE projection of tweets about the most shared news articles.}
    \label{fig:tsne}
\end{figure}

Where $\bm{f_{t'}}$ is an irrelevant tweet mined using the multi-similarity method \cite{wang2019multisimilarity}, and $\alpha$ is the margin. This minimizes the distance between the embeddings of articles and their relevant tweet(s). Figure \ref{fig:tsne} shows the t-Stochastic Neighbor Embedding (t-SNE) \cite{maaten2008visualizing} projection of the learned embeddings for a sample of tweets which reference the top-10 most shared articles in the dataset. The result is that tweets referring to the same articles are tightly clustered in latent space. We call our resulting model the Adaptive Siamese Transformer (\textbf{AST}).

\section{Experimental Results}
\subsection{Evaluation Criteria}
The goal of our work is to enable the automatic filtering of a corpus of tweets using news articles as seeds. We assume that it is cheap to \textit{collect} tweets, but expensive to \textit{annotate} them, and that an ideal model identifies relevant tweets while returning as close to $0$ false positives as possible. 

Precision, a common metric for many information retrieval tasks, reflects our wish to minimize the number of false positives. A model that retrieves exclusively relevant documents will have a precision of $100\%$. However, this could be achieved by simply returning a very small amount of relevant documents. We ensure that a reasonable number of documents are returned by using $r$-Precision as our evaluation metric, which computes the precision after $r$ documents are retrieved. We also report several other metrics for additional comparison. 

\subsection{Inference Process}

For evaluation, we hold out approximately $3300$ samples as our test set. The model returns an embedding for each news article $\bm{f_a}$ and tweet $\bm{f_t}$. The relevance score of a given $\bm{f_{a_i}}$ to a given $\bm{f_{t_j}}$ is defined as their cosine similarity.

\subsection{Training Configuration}
We train Siamese networks on a large dataset of tweet-article pairs \cite{guo2013linking} using three additional encoders as baselines, including Convolutional Neural Networks (CNN) \cite{Kim_2014}, Gated Recurrent Units (GRU) \cite{Cho_2014}, and Bidirectional GRUs (Bi-GRU) as benchmarks. Each network uses $300$-dimensional pre-trained \texttt{GloVe} embeddings \cite{pennington2014glove}. Triplet mining was performed using the \texttt{pytorch\_metric\_learning} library \cite{Musgrave2019}, and experiments were conducted with the \texttt{AllenNLP} library \cite{Gardner2017AllenNLP} using four NVIDIA V100 GPUs on Google Cloud Platform. All configuration details are available on GitHub.

\subsection{Results}

\begin{table}[htpb]
    \centering
    \begin{tabular}{|l|r|r|r|r|}
\hline
{} & \textbf{\textbf{CNN}} & \textbf{\textbf{GRU}} & \textbf{\textbf{Bi-GRU}} & \textbf{\textbf{AST}} \\ \hline
50 & \textbf{100.0} & 98.00 & 98.00 & \textbf{100.0} \\ \hline
100 & \textbf{99.00} & 96.00 & \textbf{99.00} & 96.00 \\ \hline
200 & \textbf{97.50} & 92.00 & 96.50 & 94.50 \\ \hline
500 & 90.40 & 88.00 & \textbf{92.60} & 92.40 \\ \hline
1000 & \textbf{84.80} & 79.80 & 83.70 & \textbf{84.80} \\ \hline
2000 & 68.65 & 62.25 & 67.25 & \textbf{70.00} \\ \hline
3000 & 55.53 & 49.93 & 54.27 & \textbf{57.90} \\ \hline
\end{tabular}
    \caption{$r$-Precision for each model at various thresholds.}
    \label{tab:r_precision}
\end{table}

\paragraph{Evaluation \#1.} We report the $r$-Precision for each baseline encoder, which is the precision after having retrieved $r$ documents. For each $r$, we retrieve the $r$ highest scoring tweet-article pairs by cosine similarity, and compute the precision. This captures both the quality and quantity of documents retrieved by each model. All models decrease in precision as they retrieve more documents, as is expected \cite{manning1999foundations}. The GRU-based framework quickly falters in performance, demonstrating the challenges recurrent models face with long sequences. Bi-directionality appears to help, but the most competitive baseline was clearly the CNN-based framework. We posit that CNNs' ability to identify the most important n-grams using global max pooling \cite{jacovi-etal-2018-understanding} provides the filtering ability we describe in our first objective. Our model performs the best at higher quantity thresholds, significantly out-performing the recurrent models. It is possible that the sliding-window approach of the Star Transformer could lead to similar behavior as CNNs, a possibility we leave to be examined in future work.

\begin{table}[htpb]
\centering
\begin{tabular}{|c|c|c|c|}
\hline
 & \textbf{mAP} & \textbf{AUC-ROC} & \textbf{mRP} \\ \hline
\textbf{GRU} & 43.53 & 96.03 & 76.68 \\ \hline
\textbf{Bi-GRU} & 47.58 & 96.05 & 80.29 \\ \hline
\textbf{CNN} & 49.81 & 97.09 & 81.05 \\ \hline
\textbf{AST} & \textbf{53.01} & \textbf{97.50} & \textbf{81.31} \\ \hline
\end{tabular}
\caption{Comparison of mean Average Precision, area under the Receiver-Operating Curve, and mean $r$-Precision between all models.}
\label{tab:my-table}
\end{table}

\paragraph{Evaluation \#2.} We also report \textbf{m}ean  $\mathbf{r}$-\textbf{P}recision (\textbf{mRP}), Mean average precision (\textbf{mAP}), and area under the Receiver-Operating Curve (\textbf{AUC-ROC}). \textbf{mRP} is simply the $r$-Precision over all thresholds in Table \ref{tab:r_precision}. \textbf{mAP} and \textbf{AUC-ROC} are calculated using the \texttt{scikit-learn} library \cite{scikit-learn}. Our model outperforms the three baselines in each metric.

\begin{figure}
    \centering
    \includegraphics[width=0.8\linewidth]{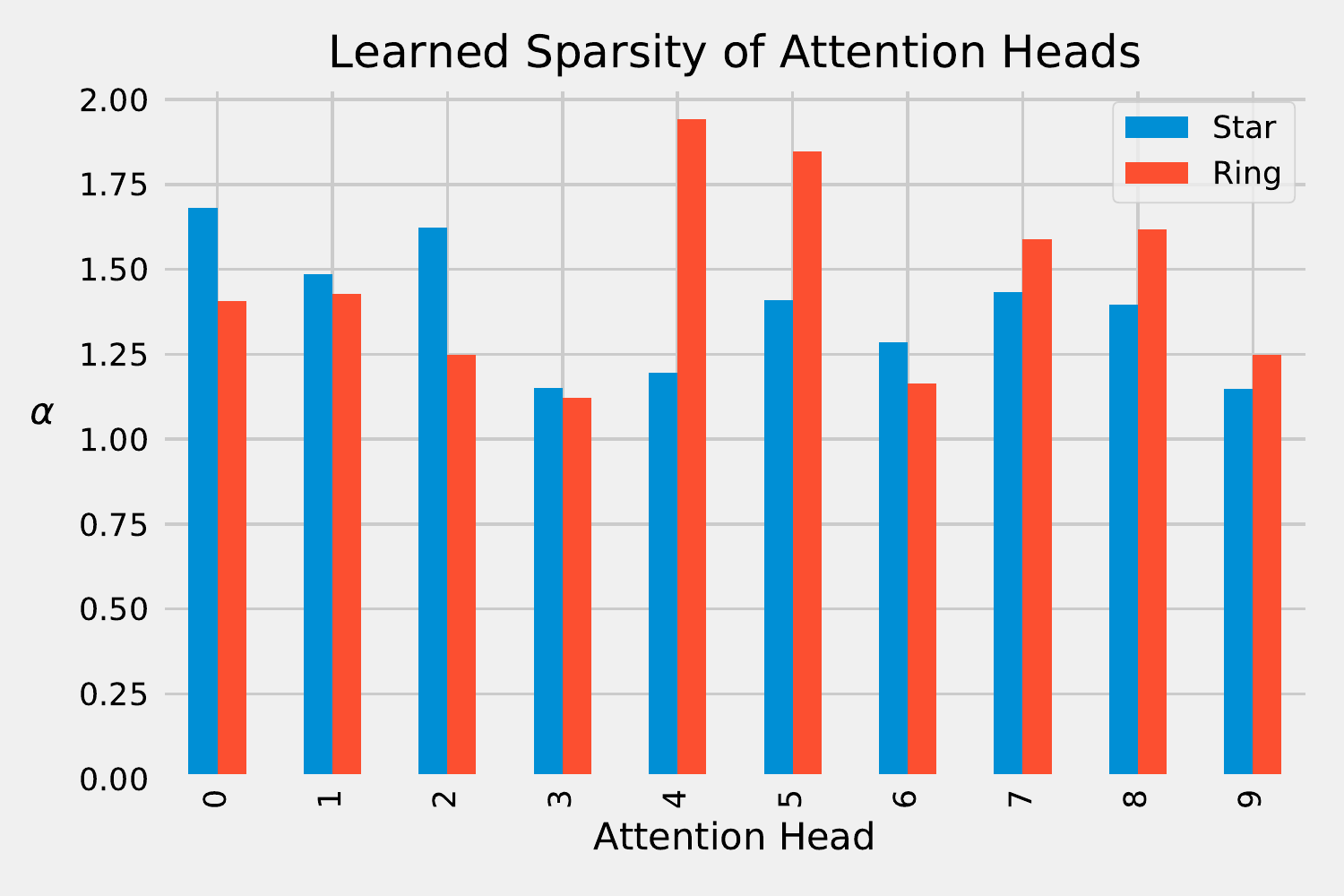}
    \caption{Learned $\bm{\alpha}$ for each attention head.}
    \label{fig:my_label}
\end{figure}

\paragraph{Learned sparsity.} Following the training process, we examine the $\bm{\alpha}$ values learned by each attention head of the Star Transformer. Values closer to $1$ will lead to attention weights closer to those produced by $\softmax$, while values closer to $2$ will be similar to sparsemax, and a value of $1.5$ is equal to $1.5-\mathrm{entmax}$ \cite{Correia_2019}. The star attention heads are used to update the relay node, while the ring attention heads are used at each context window. Overall, it appears that the ring attention heads learn a diverse set of alphas, allowing for varying levels of sparsity in the representations of each context window. Heads $4$ and $5$ approach sparsemax, while heads $3$ and $6$ remain relatively close to $\softmax$. The $\bm{\alpha}$ values for the star attention heads were more tightly bound, but heads $1$ and $3$ were still quite sparse, exceeding the sparsity of $1.5-\mathrm{entmax}$.

\section{Case Study}
\begin{figure}[ht]
    \centering
    \includegraphics[width=0.6\linewidth]{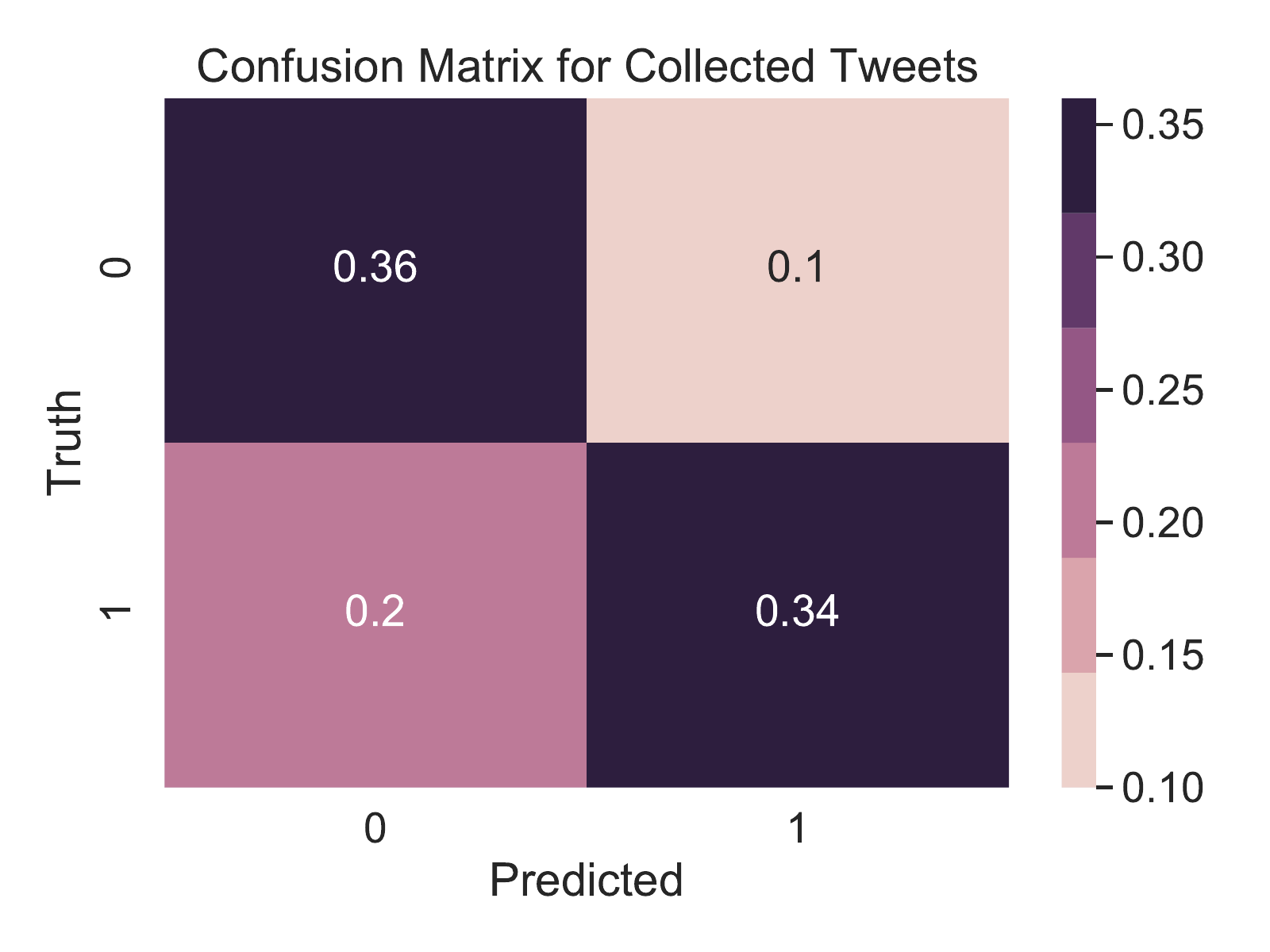}
    \caption{Of all tweets retrieved by our model, 36 were relevant while only 10 were irrelevant.}
    \label{fig:cm}
\end{figure}

Finally, we present a case study applying our model to a corpus of tweets about a terrorist attack which took place in Toronto, Canada in April 2018. We computed the cosine similarity of 100 manually annotated tweets' embeddings to that of a news article written shortly after the attack\footnote{\url{https://globalnews.ca/news/4161785/pedestrians-hit-white-van-toronto/}}. As shown in Figure \ref{fig:cm}, with a retrieval threshold of $\cos(\bm{f_a, f_t})\geq 0.7$, our model was able to minimize the number of false negatives without greatly diminishing the quantity of tweets retrieved.

As seen in Table \ref{tab:casestudy}, sorting the tweets by cosine similarity showed results which may suggest that the model was able to learn which tweets are more relevant than others (as opposed to a binary notion of relevance), though we note these results are heavily anecdotal. We share these results to demonstrate potentially useful applications of our model. For example, a project which requires at least $N$ relevant tweets and precisely $0$ irrelevant tweets necessarily requires human annotations. Our model could significantly decrease the financial cost of collecting annotations by presenting the tweets in order of estimated relevance (cosine similarity), minimizing the quantity of tweets that must be shown to annotators until $N$ relevant tweets are found.

\begin{table}[htpb]
    \centering
    \begin{tabular}{p{12cm}p{1cm}}
\toprule
\textbf{tweet} & \textbf{score} \\
\midrule
 Second, my thoughts and prayers go out to everyone in Toronto after the attack that left 10 dead and 16 injured! A d also, thank you so much for a... &  0.9166 \\
     What a devastating Tragedy! Savagery of mankind has no bound! Deadly Toronto van driver: What we know about Alek Minassian  http://bit.ly/2vFJfDx &  0.8996 \\
 Everyone at Community Safety Net would like to express their deepest sympathies to all those affected by yesterday's events in Toronto \#TorontoStr... &  0.8841 \\
                                             What we know about Alek Minassian, the man charged in deadly Toronto van attack  http://ow.ly/jZpd30jF9Jo &  0.8838 \\
 Yesterday was a deeply sad day for the city of \#Toronto and our hearts go out to the victims, the families and anyone affected by the tragic incid... &  0.8686 \\\\
  \multicolumn{1}{c}{\vdots}   & \multicolumn{1}{c}{\vdots}      \\\\
                                        Incel or MGTOW Facebook groups banned me from their Facebook groups all because they also hate not just women. &  0.4311 \\
                                                                                                          Une pense pour le tragique vnement  Toronto  &  0.3877 \\
                                                                                                                 Stay strong Canada - we are with you  &  0.3875 \\
 It is both wrong and unhelpful to cast the incel phenomenon as some kind of dark internet cult. This is garden variety misogyny and rape culture w... &  0.3797 \\
 Reading that InCel "female ratings card" post going around and am increasingly beset by a burning curiosity as to where lesbians fit into their co... &  0.3361 \\
\bottomrule
\end{tabular}
\caption{The most and least similar Tweets to a news article written after the attack.}
\label{tab:casestudy}
\end{table}

\section{Discussion}
In conclusion, we present a novel approach to article-comment linking using a Siamese architecture and triplet loss. We encode pairs using Adaptive Star Transformers, an efficient Transformer using adaptively sparse attention to filter irrelevant information from the input sequences, which we show outperforms several other encoders in a Siamese framework. This model could allow other researchers to eliminate or vastly reduce the cost of filtering corpora.

\section{Acknowledgements}
Thank you to Kevin Musgrave for assistance using his \texttt{pytorch\_metric\_learning} library, and to Jorge Salazar for consultation on the proof regarding the softmax function and general proof-reading. 

\newpage
\bibliographystyle{plain}
\bibliography{neurips_2019}

\newpage
\begin{appendices}

\section{Softmax Expectation}
\label{appendix:expsum}
We will show that an increase in $S$, the length of a document, can be expected to diminish the weight of each token in the softmax step of the attention mechanism. To demonstrate a decrease in $\ev{\sigma(\bm{z})_i}$ with $S$, we need only show an increase in the expectation on the denominator $\sum^{S}_{j=1} e^{z_j}$, which we refer to as $\expsum$.

Using cosine attention, let $\bm{z} = \cos(\bm{Q, K})$. Since $\cos(\cdot) \in [-1, 1]$, let each $z_j$ be an independent random variable drawn uniformly from $\mathcal{U}(-1, 1)$. It follows that all $e^{z_j} \in [e^{-1}, e^1]$. The expectation on $\expsum$ then becomes:


\begin{align}
    \ev{\sum^{S}_{j=1} e^{z_j}} &= \sum^{S}_{j=1} \ev{e^{z_j}} \\
    &= \sum^{S}_{j=1} \int_{-1}^1 \frac{1}{2}e^{z_j} \\
    &= \sum^{S}_{j=1} \frac{e^2 - 1}{2e} \\
    &= S\cdot\frac{e^2 - 1}{2e} \\
    &\approx 1.1752\cdot S
\end{align}

This demonstrates that $\ev{\expsum}$ increases as a function of $S$. As stated previously, an increase in $\ev{\expsum}$ decreases  $\sigma(\bm{z})_i$. This dictates that $\sigma(\bm{z})_i$ must decrease when $S$ increases. Therefore, an increase in the length of a document will decrease the expected attention weight of each token. $\blacksquare$

\end{appendices}

\end{document}